\documentclass[aps,prl,twocolumn,showpacs,floatfix,superscriptaddress,amssymb,
amsmath,preprintnumbers]{revtex4}
\usepackage{graphicx}
\usepackage{dcolumn}
\usepackage{bm}
\usepackage{amsmath}

\usepackage{color}

\begin{document}
\title{Kondo screening suppression by spin-orbit interaction in quantum dots}
\author{E. Vernek}
\affiliation{Department of Physics and Astronomy, and Nanoscale
and Quantum Phenomena Institute, \\Ohio University, Athens, Ohio
45701-2979} 
\affiliation{Instituto de F\'isica, Universidade Federal de Uberl\^andia, Uberl\^andia, 38400-902, MG - Brazil } 
\affiliation{Kavli Institute for Theoretical Physics, University of California, 
Santa Barbara, CA 93106-4030}

\author{N. Sandler}
\affiliation{Department of Physics and Astronomy, and Nanoscale
and Quantum Phenomena Institute, \\Ohio University, Athens, Ohio
45701-2979}
\affiliation{Kavli Institute for Theoretical Physics, University of California, 
Santa Barbara, CA 93106-4030}

\author{S. E. Ulloa}
\affiliation{Department of Physics and Astronomy, and Nanoscale
and Quantum Phenomena Institute, \\Ohio University, Athens, Ohio
45701-2979}
\affiliation{Kavli Institute for Theoretical Physics, University of California, 
Santa Barbara, CA 93106-4030}

\date{\today}
\preprint{NSF-KITP-09-15}

\begin{abstract}
We study the transport properties of a quantum dot embedded
in an Aharonov-Bohm ring in the presence of spin-orbit interactions.  Using a numerical renormalization group analysis of the system in
the Kondo regime, we find that the competition of
Aharonov-Bohm and spin-orbit dynamical phases induces a strong suppression of the Kondo state singlet, somewhat akin to an effective intrinsic magnetic field in the system.  This effective field breaks the spin degeneracy of the localized state and produces a finite magnetic moment in the dot. By introducing an {\em in-plane} Zeeman field we show that the Kondo resonance can be fully restored, reestablishing the spin singlet and a desired spin filtering behavior in the Kondo regime, which may result in full spin polarization of the current through the ring. 
\end{abstract} 
 
 
\maketitle

Semiconductor quantum dot (QD) structures represent a promising platform on which to achieve charge and spin control, due to their discrete energy levels, sizable Coulomb interaction due to strong electron confinement, as well as precise level and size manipulation via gate voltages \cite{QDs}. This flexibility allows probing fundamental aspects of spin systems and opens possibilities for devices with newly tailored properties \cite{Spintronics}.  Interestingly, coherent electron propagation at low temperatures and quantum interference may play a pivotal role in attaining the desired goal in these structures \cite{Grbic}.

Control of electronic transport is now systematically achieved by exploiting interference in multiple-path geometries, such as those produced with one or multiple QDs embedded in a ring \cite{Clerk-Fuhrer}. In these structures, a weak magnetic field through a ring ($\simeq$few mT for a submicron ring) produces significant changes in transport properties due to the Aharonov-Bohm (AB) effect \cite{Aharonov-Bohm}.  The presence of Rashba spin-orbit (SO)
interaction \cite{Rashba}, which can be further modulated by applied gate voltages, provides additional dynamical control of charge and spin transport.   
The strong Coulomb interaction in these systems may also result in a Kondo state 
appearing below a characteristic Kondo temperature $T_K$ \cite{Inoshita}, giving rise to 
strong antiferromagnetic correlations between localized and itinerant electrons in the leads.  The
Kondo state singlet produces an additional transport channel through the system at the Fermi level when the dot is in an otherwise Coulomb blockaded configuration.  

Several ring geometries have been proposed as {\em spin polarizers} and their behavior has been analyzed in different regimes and levels of approximation in models that include SO interactions \cite{Qing,Heary,Lobos}. The basic physics involved in the spin-filtering effect is the modification of the conductance for different spin species as a result of the AB flux and SO effects that introduce {\em spin-dependent} dynamical phases for the electrons in the multiply-connected geometry.  The correct and complete inclusion of 
the Kondo physics is crucial in order to provide a proper description of the system, as we describe in this work, especially with respect to its spin transport behavior.  One important element of this analysis is the full consideration of particle-hole (p-h) asymmetry, which has been ignored or included only approximately in previous works, and it is found to have dramatic effects in the correlations of the Kondo state for even the simplest geometry of a QD embedded in the ring.  We present here a numerical renormalization group study of this system.  This approach is capable of addressing the full spin-dependent character of the coupling to the leads and the p-h asymmetry in a non-perturbative fashion \cite{NRG}.  Our analysis demonstrates that the combination of AB and SO effects may strongly suppress the Kondo state, and in fact eliminate the desired spin filtering effect described previously in the literature \cite{Heary}.  Moreover, we demonstrate that this suppression can be fully compensated by the application of an {\em in-plane} Zeeman field.  Under those conditions, the Kondo screening is restored and the spin filtering effect reestablished.  

\begin{figure}
\centerline{\resizebox{1.8in}{1.1in}{\includegraphics{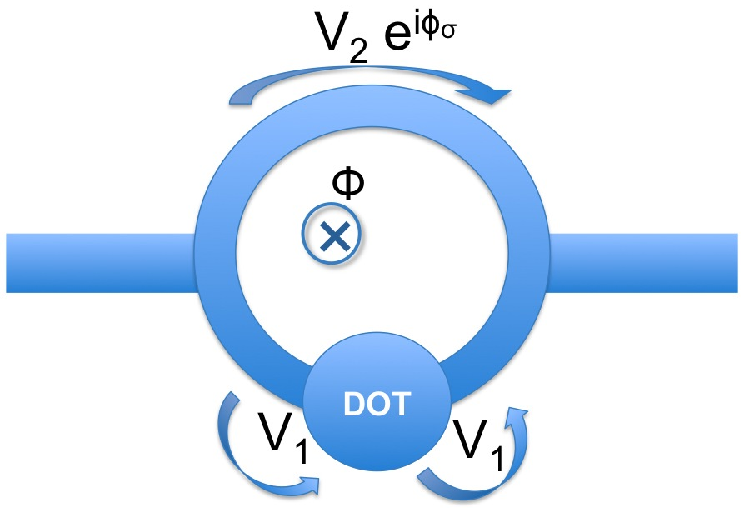}}}
\centerline{\resizebox{3.5in}{!}{\includegraphics{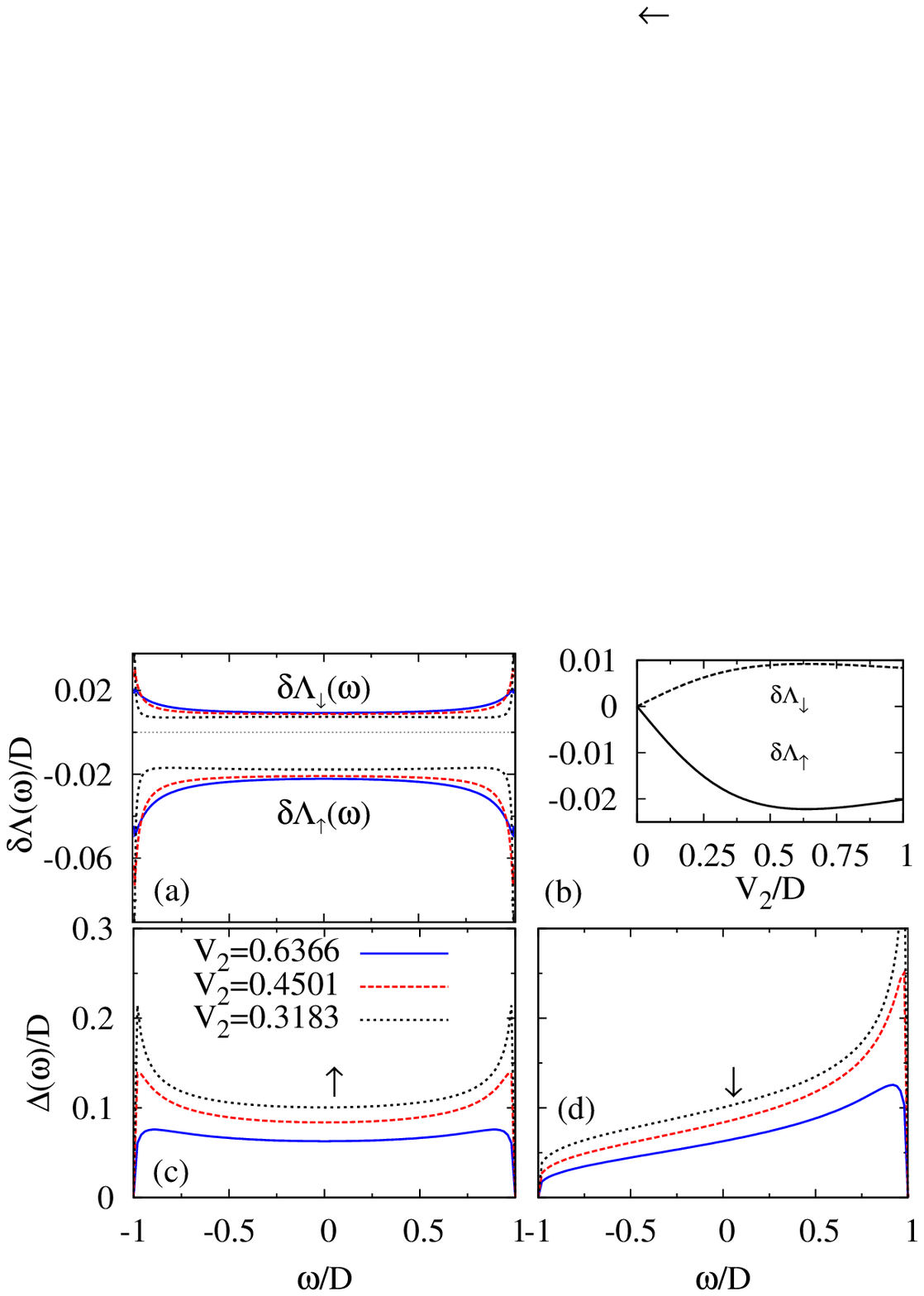}}}
\caption{\label{fig2} (color online) Top: Quantum dot embedded in Aharonov-Bohm ring. (a) Non-interacting electron self-energy vs frequency for $V_1=0.1414D$, $\phi_{AB}=\phi_{SO}=\pi/4$ and various
coupling $V_2/D$, as indicated in panel (c). (b) Self-energy energy shift $\delta \Lambda_\sigma (\omega=0)$ vs $V_2$. (c) and (d) Effective coupling to leads $\Delta_\sigma$ vs  $\omega$ for spin $\uparrow$ and $\downarrow$, respectively. }
\vspace{-1em}
\end{figure}

We consider a single QD in contact with two leads $L$
and $R$. The leads are coupled to each other via
the upper arm of the ring, as shown in the upper diagram in Fig.\ \ref{fig2}, while the lower arm contains the QD.\@
The hamiltonian of the system can be written as $H=H_{QD}+H_{leads}+H_T$, where
$H_{QD}=\sum_{\sigma}\epsilon_dn_{d\sigma}+Un_{d\uparrow}n_{d\downarrow}$ describes interacting electrons confined in the QD with level energy
$\epsilon_d$ regulated by a local gate, and  
$U$ is the local Coulomb repulsion in the dot;
$H_{leads}=\sum_{\alpha k\sigma}[c^\dagger_{\alpha
k\sigma}c_{\alpha k\sigma}+H.c.] $, with $\alpha=R,L$, describes the leads,
and $ H_{T}=\sum_{\alpha k\sigma}V_1[c^\dagger_{d\sigma}c_{\alpha
k\sigma}+H.c.]+
\sum_{\sigma}[\tilde V^\sigma_2c^\dagger_{Lk\sigma}c_{Rk\sigma} + H.c.]$,
describes the connection between the leads through both arms of the ring, 
where $c^\dagger_{d\sigma}$ creates an electron in the dot, while 
$c^\dagger_{\alpha k\sigma}$ creates it in
the $\alpha$-th lead with spin $\sigma$. The coupling $V_1$ is considered real and allows
the QD electron to tunnel to/from the leads. The AB field is mapped, as
usual, into a phase $\phi_{AB}$ ($=2\pi\Phi/\Phi_0$, where $\Phi$ is the magnetic
flux through the plane of the ring, $\Phi_0=hc/e$ is the flux quantum) accumulated
when the electron undergoes a closed trajectory in the ring, so that 
$\tilde V_2 \rightarrow V_2 e^{i\phi_{AB}}$ is a complex number. 
In addition, a local SO interaction on a single-level QD can be mapped
onto a {\em spin-dependent} phase $\sigma\phi_{SO}$, which is proportional to the strength of the SO interaction and accumulates along the ring (spin quantization axis of $\sigma$ is {\em in the plane} of the ring) \cite{Qing}. 
The combined AB and SO effects can then
be included by a spin-dependent phase of the tunneling coupling $\tilde
V^\sigma_2=V_2e^{i\phi_{\sigma}}$, where $V_2=|\tilde V^\sigma_2|$ and
$\phi_\sigma=\phi_{AB}+\sigma\phi_{SO}$. The appearance of $\phi_{SO}$
in the upper arm of the ring can also be thought to arise from a variable relative gate potential applied on that arm \cite{Grbic}.

{\it Non-interacting case. ---} To study  the consequences of SO coupling and AB flux in the system, we consider first the non-interacting
limit ($U=0$). The exact Green's functions of
the system can be calculated; in particular the local GF at the QD is
$G^{(0)\sigma}_{dd}(\omega) = [\omega-\epsilon_d+\Sigma_\sigma(\omega)]^{-1}$, where the self-energy is
 \begin{eqnarray}
\Sigma_\sigma(\omega)&=&\frac{-2V_1^2}{1-V_2^2 (\tilde{G}(\omega))^2} 
\left[\tilde{G}(\omega)+V_2 (\tilde{G}(\omega))^2 \cos\phi_\sigma\right], \ \ \ \ \ 
\label{Sigma}
 \end{eqnarray}
 with $\tilde{G}(\omega)= \sum_k (\omega-\epsilon_k)^{-1}$.
It is useful to write the self-energy in terms of its real and imaginary parts, 
$\Sigma_\sigma(\omega)=\Lambda_\sigma(\omega)+i\Delta_\sigma(\omega)$.  All the information about SO and AB phases is contained in the second term of Eq.\ (\ref{Sigma}). The spin-dependent contribution to the self-energy is less important when the condition  $V_2\cos(\phi_\sigma)\ll 2D/\pi$ is fulfilled, where $D$ is the half-bandwidth of the leads \cite{Heary}. 
In the non-interacting case, even when $V_2\cos(\phi_\sigma)\approx 2D/\pi$, the spin dependent term in the self-energy is a relatively small correction. However, these changes are crucial in the interacting case, as we will discuss below. 

Notice in Eq.\ (\ref{Sigma}) that for $\phi_{AB}=0$, 
$\Sigma_{\uparrow}(\omega)=\Sigma_{\downarrow}(\omega)$ (since the cos is even) but the equality  does not hold for arbitrary values of $\phi_{AB}$.
As a consequence, one has in general $\Lambda_{\uparrow}(\omega)\neq\Lambda_{\downarrow}(\omega)$,
which means that the local bare level acquires a spin dependent  shift. This is similar to what happens when electrons are subject to a Zeeman field, although
here the shifts are different in magnitude for each spin species, and the shift is 
$\omega$-dependent.  
Experiments where an intrinsic magnetic field is observed in the single-particle spectrum have been reported recently in an AB ring with strong Rashba SO interaction \cite{Grbic}. 

To explore how this effect depends on system parameters, we define 
$\delta\Lambda_\sigma(\omega)=\Lambda_\sigma(\omega,\phi_{SO}
)-\Lambda_\sigma(\omega,0)$; this quantity plays the role of the $\sigma$-dependent magnetic field producing the spin splitting. 
Figure \ref{fig2}a shows typical curves $\delta\Lambda_\sigma(\omega)$ for $V_1=0.1414D$, $\phi_{AB}=\phi_{SO}=\pi/4$
and different values of $V_2$.  Figure \ref{fig2}b
shows $\delta\Lambda_\sigma(0)$ as function of the coupling $V_2$. The
maximum absolute value of $\delta\Lambda_\sigma(0)$ is obtained
for $V_2=2D/\pi$.  Notice that $|\delta\Lambda_\uparrow(0)|\neq
|\delta\Lambda_\downarrow(0)|$ highlights the p-h asymmetry and makes this phenomenon very different from that of an external Zeeman field. 
Figure \ref{fig2}c,d show the effective coupling $\Delta_\sigma$ that the localized
electrons in the QD have with the conducting leads.  The
SO interaction and AB effect result in drastically different couplings $\Delta_\uparrow(\omega)$ and $\Delta_\downarrow(\omega)$ [although identical at the Fermi energy, $\Delta_\uparrow (0) = \Delta_\downarrow (0)$]. This has an important effect on the interacting case, especially when the system enters the Kondo regime.  We emphasize that larger $V_2$ values correspond to better connectivity of the upper arm in the ring, while larger $\phi_{SO}$ arises from larger SO coupling \cite{Grbic}.

{\it Interacting case. ---} In the following, we take $U=0.5D$ and $\epsilon_d=-U/2$.
In the NRG approach, the system is mapped into an Anderson impurity coupled to an effective non-constant conduction band given by the effective spin-dependent coupling 
$\Delta_\sigma(\omega)={\tt Im}[\Sigma_\sigma(\omega)]$.
A generalization of the NRG discretization for a non-constant conduction
band \cite{Jayaprakash} is used to calculate the local interacting Green's function \cite{Bulla1}.  The latter is written as 
$[G^{\sigma}_{dd}(\omega)]^{-1}=\omega-\epsilon_d+\Sigma^*_\sigma(\omega)$, where 
$\Sigma^*_\sigma(\omega)$ is the proper self-energy, from which we obtain the normalized local density of states (LDOS),
defined as $\tilde{\rho}_\sigma = - \Delta_\sigma (0){\tt Im}[G^\sigma_{dd}(\omega)]$.

\begin{figure}
\begin{center}
 \resizebox{3.5in}{!}{\includegraphics{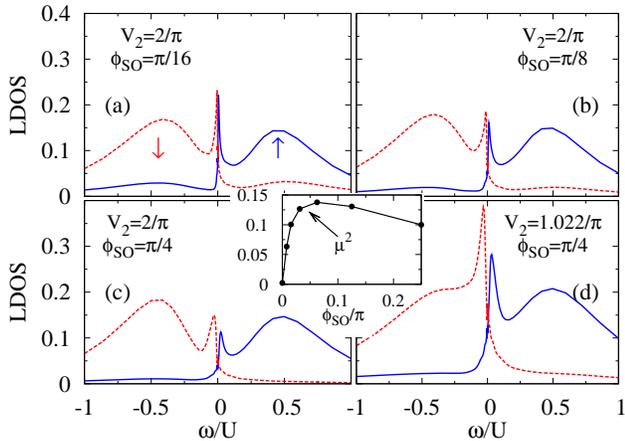}}
\caption{\label{fig3} (color online) Kondo peak splitting in the LDOS for 
$V_1=0.1414D$, $\phi_{AB}=\pi/4$, and $V_2$ and
$\phi_{SO}$ as shown. Continuous (blue) and dashed (red) lines are for spin $\uparrow$ and $\downarrow$, respectively. Central inset shows QD effective free moment $\mu^2$ for $V_2=2D/\pi$ vs $\phi_{SO}$.}
\label{test4}
\end{center}
\vspace{-2em}
\end{figure}

Figure \ref{fig3} shows the LDOS for $V_1=0.1414D$, $\phi_{AB}=\pi/4$ and
different $V_2$ and $\phi_{SO}$. Starting with $V_2=2D/\pi$
and $\phi_{SO}=\pi/16$ in Fig.\ \ref{fig3}a, one can see a small spin-resolved shift in the Kondo peak, in addition to a clear asymmetry in $\omega$ \cite{Andreas}. 
Increasing $\phi_{SO}$ in panels b and c results in stronger $\omega$-asymmetry and in weaker Kondo peaks near $\omega \simeq 0$.
Similar asymmetries have been found in a QD coupled to ferromagnetic leads \cite{Choi}, produced, as it is the case here, by the different effective couplings through which different spins in the QD couple to the conduction electrons.
The importance of the p-h asymmetry in the effective conduction band has been demonstrated before \cite{Choi}; however, here the shape of the effective coupling is determined by the AB and SO phases and not from ferromagnetism in the leads. 
In Fig.\ \ref{fig3}d we keep $\phi_{SO}=\pi/4$, as in Fig.\ \ref{fig3}c, but
decrease $V_2$ to $1.022D/\pi$;  the Kondo peak is progressively restored as  $V_2$ decreases. This behavior can be qualitatively understood in terms of the
enhanced effective coupling $\Delta_\sigma (\omega)$ (and corresponding larger $T_K$) as  $V_2$ decreases, as seen in Fig.\ \ref{fig2}c and d.  The inset in Fig.\ \ref{fig3} shows the 
QD contribution to the free magnetic moment [defined as $\mu^2=\langle S_z^2 \rangle_{\rm QD}$, where $\langle \cdots \rangle_{\rm QD}$ is the average over the ground state after subtraction of the band contribution \cite{NRG}] as function of $\phi_{SO}$ for $\phi_{AB}=\pi/4$ and $V_2=2D/\pi$. 
Notice that $\mu^2$ increases with $\phi_{SO}$, which means that the QD magnetic moment is rapidly unscreened by SO interactions. 

\begin{figure}
\begin{center}
 \resizebox{3.5in}{!}{\includegraphics{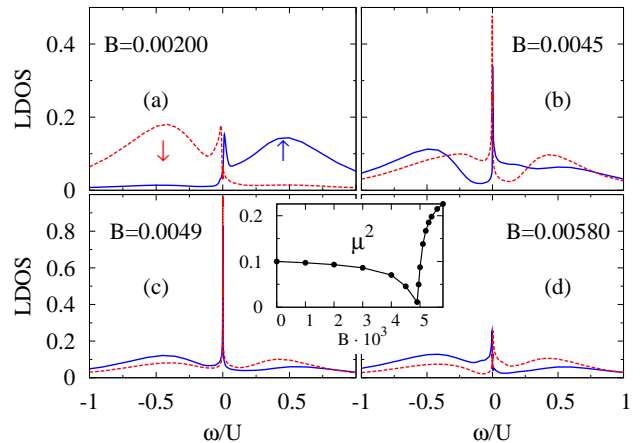}}
\caption{\label{fig4} (color online)  LDOS vs $\omega$ for Fig.\ \ref{fig3}c parameters and Zeeman field $B$: (a) $B=0.0020$, 
(b) $B=0.0045$, (c) $B=0.0049$,
and (d) $B=0.0058$. Continuous (blue) and dashed (red) lines are for spin 
$\uparrow$ and $\downarrow$, respectively.  Inset shows local magnetic moment vs field, vanishing at $B\simeq 0.0049$.}
\label{test4}
\end{center}
\vspace{-2em}
\end{figure}

As described previously \cite{Heary}, the suppression of the Kondo peak at
the Fermi level is detrimental to the spin-filtering properties of the system in this regime, as the
suppression prevents the additional path to produce the required quantum interference in the ring.  We find that a way to restore the Kondo peak suppressed by SO is to
apply an {\em in-plane} magnetic field that produces a Zeeman
shift in the QD levels \cite{Choi}.
Figure \ref{fig4} shows the LDOS  for different values of the in-plane magnetic field $B$ for the case with $V_1=0.1414D$, $V_2=2D/\pi$, and $\phi_{SO}=\phi_{AB}=\pi/4$ (same as in Fig.\ \ref{fig3}c). Notice
that by increasing the Zeeman field the Kondo peaks are progressively  restored and reach their maximum height for $B=0.0049$ (notice this field is measured in energy units, and would correspond to several tesla in GaAs---of course, the compensating field depends on the parameters of the system, especially $V_2$ and 
$\phi_{SO}$).  Increasing $B$ further, rapidly suppresses the Kondo peak, as seen in  Fig.\ \ref{fig4}d, overcompensating the intrinsic magnetic field. The inset of Fig.\ \ref{fig4} depicts $\mu^2$ vs $B$, showing that 
$\mu^2\rightarrow 0$ as $B\rightarrow 0.0049$, and the complete screening of the local magnetic moment is restored.  Notice that the amplitude of the Kondo peak, although restored by the Zeeman field, is somewhat different for spin $\uparrow$ and $\downarrow$. 
This asymmetry in spins arises not only from differences in $\Delta_\sigma$, especially as $\Delta_\uparrow (0) = \Delta_\downarrow (0)$, but rather from the fact that $|\delta\Lambda_\uparrow (0) | > \delta\Lambda_\downarrow (0)$, as seen in Fig.\ \ref{fig2}b.  This demonstrates the non-trivial effect of the p-h asymmetry and the importance of considering it fully when evaluating the role of electronic interactions.  Moreover, as we will now describe, this asymmetry has dramatic effects on the conductance of the system and on its spin filtering properties.  

We calculate the conductance in the zero-bias regime, which can be  written as \cite{Hofstetter,Heary},
\begin{eqnarray}\label{cond}
 G_\sigma/G_0&=&T_0-2\tilde\Delta_0\sqrt{T_0R_0}\cos(\phi_\sigma){\tt
Re}\,G^\sigma_{dd}(0)\nonumber\\
&-&{\tilde\Delta_0}\left\{1-T_0[1+\cos^2(\phi_\sigma)]\right\}{\tt
Im}\,G^\sigma_{dd}(0)  ,
\end{eqnarray}
where $G_0=e^2/h$, $T_0=4r/(1+r)^2$ is the transmission through the upper arm
of the ring, $R_0=1-T_0$, $\tilde\Delta_0=\Delta_0/(1+r)$, with
$r=\pi^2V_2^2/4D^2$, and $\Delta_0=2\pi V_1^2/D$.

\begin{figure}
\begin{center}
\resizebox{3.25in}{!}{\includegraphics{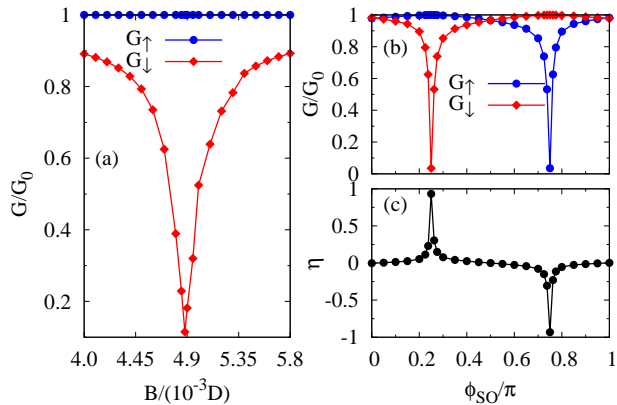}}
\caption{\label{fig5} (color online) (a) Conductance vs Zeeman field for Fig.\ \ref{fig3}c parameters. (b) $G_\sigma$ vs SO phase for 
$V_1=0.1414D$, $V_2=2D/\pi$, $\phi_{AB}=\pi/4$ and $B=0.0049$.
(c) Conductance polarization $\eta$
(see text) vs $\phi_{SO}$ for conductances of panel (b). }
\end{center}
\vspace{-2em}
\end{figure}

The conductance $G_\sigma$ is shown in Fig.\ \ref{fig5}a as function of the Zeeman field, for the same
parameters of Fig.\ \ref{fig4}. Notice that while $G_\uparrow$
remains in the unitary limit, $G_\downarrow$ drops sharply near 
$B\approx 0.0049$. This behavior can be understood in terms
of the contribution to the
conductance in Eq.\ (\ref{cond}): For $V_2=2D/\pi, $ $r=T_0=1$ and
$R_0=0$; in this case the second term gives no contribution
and the third term becomes $\tilde\Delta_0\cos^2(\phi_\sigma){\tt
Im}[G^\sigma_{dd}(0)]$. For $\phi_{SO} \simeq \phi_{AB}=\pi/4$,
$G_\uparrow/G_0\approx1$  and $G_\downarrow/G_0=1-
\tilde\Delta_0\pi\rho_\downarrow(0)\approx 0$ \cite{Friedel}, which gives the spin
filtering condition (other values of the phases result in smaller contrast for the two spins). In Fig.\ \ref{fig5}b we fix $\phi_{AB}=\pi/4$ and $B=0.0049$ and
plot the conductance as function of $\phi_{SO}$. We see that $G_\uparrow$
remains close to the unitary limit in the interval $0<\phi_{SO}<\pi/2$
while $G_\downarrow$ vanishes for $\phi_{SO}=\pi/4$; the opposite
occurs in the complementary interval. The spin polarization of the conductance, $\eta=(G_\uparrow-G_\downarrow)/(G_\uparrow+G_\downarrow)$,
 as function of $\phi_{SO}$ is shown in Fig.\ \ref{fig5}c. Notice that for
$\phi_{SO}=\pi/4$ and $\phi_{SO}=3\pi/4$ the system exhibits almost perfect spin
polarized conductance. 

In conclusion, we have shown that the combination of 
SO interaction and AB effect results in an effective magnetic field that strongly suppresses the Kondo resonance. As a consequence, the transport behavior and the
possibility of producing spin polarized conductance is greatly affected.
However, we show it is possible to fully restore the Kondo state screening and spin polarized transport by applying an in-plane
Zeeman field.  Apart from its importance in spin polarized transport, this effect emphasizes the subtle interplay of many body correlations and their control via external fields.  It would be interesting to explore this effect by measuring spin polarized currents in this geometry.

We acknowledge helpful discussions with K. Ingersent, C. B\"usser, and L. Dias da Silva, the hospitality of the KITP, and the support of NSF-MWN-CIAM 0710581, NSF PHY05-51164, and OU BNNT.  


\begin{thebibliography}{apssamp}
\bibitem{Friedel} One can show that $\Delta_\sigma(0)=\tilde\Delta_0$, so that the Friedel sum rule for the fully compensated case yields $\rho_\downarrow(0)=1/\pi\tilde\Delta_0$.
\end{thebibliography}
\end{document}